\documentclass[twocolumn,showpacs,preprintnumbers,amsmath,amssymb,superscriptaddress]{revtex4}
\usepackage{graphicx}
\usepackage{cases} 
\usepackage{color}

\begin{document}
\title{Correlations of correlations: Secondary autocorrelations in
  finite harmonic systems} \author{Dan Plyukhin}
\email{dplyukhin@cs.utoronto.ca} \affiliation{ Department of
  Computer Science, University of Toronto, Toronto, ON, Canada}
\author{Alex V. Plyukhin} \email{aplyukhin@anselm.edu} \affiliation{
  Department of Mathematics, Saint Anselm College, Manchester, NH,
  USA}

\date{\today}% It is always \today, today,
             %  but any date may be explicitly specified

\begin{abstract}
  The momentum or velocity autocorrelation function $C(t)$ for a tagged
  oscillator in a finite harmonic system decays like that of an infinite
  system for short times, but exhibits erratic behavior at longer time
  scales.  We introduce the autocorrelation function of the long-time
  noisy tail of $C(t)$ (``a correlation of the correlation''), which
  characterizes the distribution of recurrence times.
  Remarkably, for harmonic systems with same-mass particles
  this {\it secondary} correlation may coincide with the primary correlation
  $C(t)$ (when both functions are normalized) either exactly, or over
  a significant  initial time interval. When the tagged particle is heavier than the rest, the equality does not hold, correlations shows non-random long-time scale pattern, and higher order correlations converge to the lowest normal mode. 
\end{abstract}


\maketitle

\section{Introduction}
The theme of fluctuations in finite systems of harmonic oscillators
emerges naturally in both application and theory. From a theoretical
point of view, the study of the stochastic dynamics of 
a tagged degree of freedom
in finite harmonic systems provides a valuable illustration,
and often more than that, of the role of the thermodynamic and weak-coupling limits,
ergodicity, thermalization, recurrences, synchronization, and
other basic concepts in nonequilibrium
phenomena~\cite{Zwanzig,Mazur,Cukier,Kac,Vig,Jin,Bend}.  
Another relevant area is Langevin dynamics generated by a coupling %% a coupling?
to a finite harmonic bath(s), and its application to mesoscopic systems and networks; see
\cite{PS,Hanggi,Zhou,Beims,Onofrio,Hasegawa,Akai}.

Being nonergodic, the capability of harmonic systems to illustrate general
phenomena in statistical mechanics might seem doubtful at first glance. 
By means of a canonical
transformation a harmonic system of any size can be transformed into
a collection of independent oscillators, or normal modes;
since the energies of normal modes are the integrals of motion, a single
isolated harmonic system does not equilibrate and is not very
interesting from the point of view of statistical mechanics.  

A more fruitful % and useful
approach is to consider an {\it ensemble} of
harmonic systems, assuming that in the past they were in
contact with a larger thermal bath
in equilibrium at a given temperature,
and that the initial normal modes of the ensemble are distributed
according to the canonical distribution.
Within this framework, one evaluates statistical averages of dynamical
variables over the ensemble of the system's initial coordinates rather
than over time.  Such averages show the transition 
of the ensemble to thermal
equilibrium in the limit of a large number of particles, and thus the nonergodic nature of harmonic systems does
not explicitly manifest itself, and for most cases is inessential.  It
should however be stressed that this framework, which is standard for
most works on stochastic dynamics of harmonic systems both classical
and quantum, assumes a very special type of coupling between the
system and the external thermal bath: this coupling justifies the initial
conditions for the the system's degrees of freedom, yet is assumed to
be sufficiently weak, or completely turned off, as not to affect the system's further dynamics.

The inequivalence for harmonic systems of ensemble- and time-averages,
together with the almost exclusive exploitation in literature of the
former, does not necessarily entail that the latter are inadequate.
Rather, we introduce in this paper a new class of time-average
correlations (we call these {\it secondary} correlations) 
which characterize recurrences in finite harmonic systems.
For systems of same-mass particles,
these correlations are shown to be very close, and under certain
conditions exactly identical, to the conventional (primary) time
correlations defined with ensemble averaging.
% But what do we gain by this, if they are identical?
This implies that for finite nonergodic systems, 
%even when the ensemble paradigm is adopted,
the use of both ensemble and time averages may give
meaningful complementary descriptions,
and that correlations with the two types of averaging
may be related in some subtle way.

\section{Secondary correlations}

Consider the temporal autocorrelation function
$\langle A(0) A(t)\rangle$ of a dynamical
variable $A$ in a finite system of size $L$ -
typically, such a function exhibits two distinctive regimes, separated by
a crossover time $t_c$ of order $L/v$, where $v$ is the speed of
signal propagation in the system.
For short times $t<t_c$, the variable does not feel the presence of
the boundaries, and the correlation function decays in a smooth
regular way, following the same laws as for an infinitely large system.
On the other hand, for longer times $t>t_c$ the dynamics of the
variable are affected by signals reflected from the boundaries. For
long time regimes such as this, rather than decaying smoothly the
correlation functions may exhibit erratic, apparently noisy,
behavior~\cite{Zwanzig,Mazur}.

\begin{figure}
  \includegraphics[height=6.0cm]{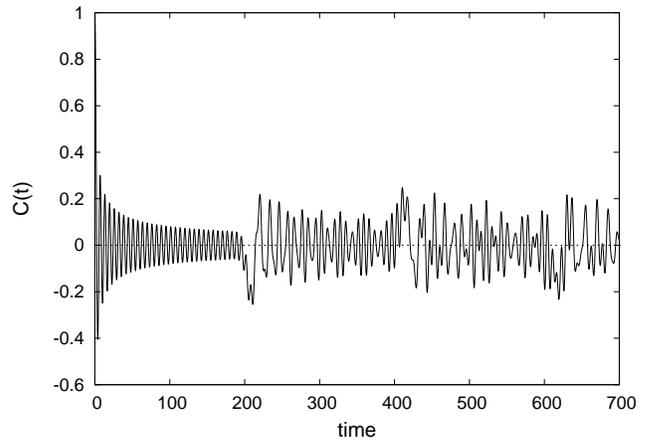}
  \caption{The normalized momentum correlation function $C_{i}(t)$ for the 
    central particle ($i=i_0=51$) of the harmonic chain with fixed ends 
    with Hamiltonian (\ref{H}) with $N=101$, 
    given by Eq. (\ref{C}). 
    The time unit is $1/2\omega$. 
The time of the crossover from the regular dissipation to 
``stochastic'' regimes is $t_c\approx 200$. %For the stochastic regime $t>t_c$
    % we evaluate the normalized secondary 
    % correlation function 
    % $D_{i_0}(t)\sim\langle C_{i_0}(\tau)\,C_{i_0}(\tau+t)\rangle_\tau$, 
    % defined by Eqs.(\ref{D1}) and (\ref{D2}). A reporting  
    % result is  that the secondary correlation $D_{i_0}(t)$, 
    % when normalized to unity at $t=0$,  coincides with the primary correlation 
    % function $D_{i_0}(t)$.
  }
  \label{fig_1}       % Give a unique label
\end{figure}

We illustrate this behavior in Fig.~\ref{fig_1} by way of the
normalized momentum correlation function for the central particle in a
harmonic chain with fixed ends. The Hamiltonian of the system is
\begin{eqnarray}
  H=\frac{1}{2m}\sum_{i=1}^N p_i^2+\frac{m\omega^2}{2}\sum_{i=0}^N(q_i-q_{i+1})^2,
  \label{H}
\end{eqnarray}
which describes $N+2$ linearly coupled particles indexed
$i=0,1,...,N+1$ with terminal particles fixed, with displacement
$q_0=q_{N+1}=0$. Assuming $N$ is odd, the middle particle indexed
\begin{eqnarray}
  i_0=\frac{N+1}{2} 
  \label{i0}
\end{eqnarray}
has normalized ($C(0)=1$) momentum correlation function
\begin{eqnarray}
\!\!\!
C_{i_0}(t)=\frac{1}{\langle p_{i_0}^2(0)\rangle}\langle p_{i_0}(0)p_{i_0}(t)\rangle
  =\frac{2}{N+1}
  \sideset{}{'}
  \sum_{j=1}^N\cos\omega_j t,
  \label{C}
\end{eqnarray}
where the prime indicates that the summation is only over odd $j$. In
this expression (we outline its derivation in the Appendix),
the $\omega_j$ terms are frequencies of normal modes
\begin{eqnarray}
  \omega_j = 2\,\omega\,\sin\frac{\pi j}{2(N+1)}
  \label{omegas}
\end{eqnarray}
where $\omega$ is the frequency of a single oscillator, and the
average $\langle \cdots\rangle$ is taken over the equilibrium ensemble
of initial conditions.  For $t<t_c$ the
correlation function $C_i(t)$ is very close to that of an infinite
chain, given by the Bessel function
\begin{eqnarray}
  C_{i}(t)\approx C_\infty(t)=\lim_{i, N\to\infty}C_i(t)=J_0(2\omega t).
  \label{limit}
\end{eqnarray} 
This can be readily justified by approximating the sum (\ref{C}) with an
integral, and recognizing the latter as the well-known
integral representation of $J_0(2\omega t)$, see e.g.~\cite{Lee,PS}.

More interesting from the perspective of this paper is the regime
$t>t_c$ in which the correlation $C_i(t)$ becomes irregular, see Fig. 1.
%reflecting recurrences of the tagged variable's values. 
% What does that mean? Seems vague.
%(These should not be
%confused with Poincar\'e recurrences, which usually return to
%the original phase point of the system (the whole chain, in our
%case) and occur on a much longer time scale.) 
It can be shown that the
function $C_i(t)$ given by (\ref{C}) belongs to the class of almost
periodic functions: any value $c$ which the function achieves once is
achieved again, infinitely many times.  Traditionally, such functions
are characterized by the average frequency with which they return to
$c$, or by the reciprocal, i.e.~the mean recurrence time $\tau(c)$.  For
correlations of type (\ref{C}) with large $N$, the famous result for
the recurrence time, first obtained by Kac~\cite{Kac} (see
also~\cite{Zwanzig,Mazur,Vig}),
\begin{eqnarray}
  \tau(c)\sim e^{Nc^2}
  \label{Kac}
\end{eqnarray}
implies that recurrences of order $c\sim N^0$ or larger are
exponentially rare. This result resolves, or rather (being derived for
a model system) shows the direction of resolution for the paradoxes of
irreversibility~\cite{Zwanzig}.

In this paper we propose to characterize the irregular part of the
function $C_i(t)$ in another way, which is more in the spirit of
nonequilibrium statistical mechanics than the mathematics of almost
periodic functions. Namely, observing that for large $t$ the
correlation function $C_i(t)$ appears to behave like stationary noise,
we are encouraged to characterize it by a new correlation function
\begin{eqnarray}
  D_i(t)=\frac{1}{\langle C_i^2(\tau)\rangle_\tau}\, 
  \langle C_i(\tau)\, C_i(\tau+t)\rangle_\tau
  \label{D1}
\end{eqnarray}
defined with the time average
\begin{eqnarray}
  \langle ...\rangle_\tau=\lim_{T\to\infty}\,\frac{1}{T}\,\int_{0}^T (...)\,d\tau.
  \label{D2}
\end{eqnarray}
Since we are only interested in the interval $t>t_c$ when $C_i(t)$ behaves irregularly, one might prefer
to set the lower integration limit in definition (\ref{D2}) to
$t_c$ instead of zero. However this would only be an unnecessary complication,
as the limit $T\to \infty$ makes the two definitions numerically
equivalent 
(assuming always that the integral from $0$ to $t_c$ converges).

We shall refer to $D_i(t)$, defined by relations
(\ref{D1}) and (\ref{D2}), as the secondary correlation function,
and call $C_i(t)$ the primary one.
% and alternatively denoted as $C(t)= C_ 1(t)$.
We would like to promote the secondary correlation $D_i(t)$ as a
meaningful statistical tool for characterizing the 
distribution of 
recurrences times in a system of finite size.  Such information is not
contained in the Kac formula~(\ref{Kac}) for the average recurrence time
$\tau$, so the two functions $\tau(c)$ and $D_i(t)$ do not
duplicate each other but describe recurrences in complementary ways.

\section{Relation to primary correlations}
Since the primary and secondary correlations $C_i(t)$ and $D_i(t)$
characterize  
recurrences at different levels and are defined using 
different types of averaging (over ensemble and time, respectively),
the existence of any specific relation between them is perhaps 
{\it a priori} unexpected. Yet a simple numerical experiment with
Eqs.~(1-6) suggests, 
for the middle atom of a chain with fixed ends, the equality 
\begin{eqnarray}
C_{i_0}(t)=D_{i_0}(t).
\label{equality}
\end{eqnarray}
Closer scrutiny reveals that the equality  
is exact and holds for any $t$, such that the 
secondary correlation completely repeats the structure of the primary one
for both regular ($t<t_c$) and noisy ($t> t_c$) domains and has the same 
crossover time $t_c$.  
The proof
follows immediately from 
the relation
\begin{eqnarray}
\langle\cos \omega_j\tau\,\cos \omega_{j'}(\tau+t)\rangle_\tau=
\frac{\delta_{jj'}}{2}\,\cos\omega_j t
\label{relation}
\end{eqnarray}
which holds for an arbitrary spectrum of (nonzero) normal mode frequencies
$\{\omega_j\}$ and can be verified by direct evaluation 
(with the help of L'Hospital's rule). 
For $t=0$ this may further be reduced to the   
familiar orthogonality relation 
for the Fourier basis, and thus can be considered
a generalized form of the latter.
From (\ref{C}) and (\ref{relation}) one obtains for the non-normalized 
secondary correlation
\begin{eqnarray}
&&\langle C_{i_0}(\tau)\,C_{i_0}(\tau+t)\rangle_\tau=\nonumber\\
&&\left(\frac{2}{N+1}\right)^2\!
\sideset{}{'}
\sum_{j,k=1}^N \langle\cos\omega_j \tau\, \cos\omega_k(\tau+t)\rangle_\tau=
\nonumber\\
&&\frac{1}{2} \left(\frac{2}{N+1}\right)^2
\sideset{}{'}
\sum_{j=1}^N \cos\omega_j t.
\label{ax1}
\end{eqnarray} 
Normalizing this function to unity at $t=0$ by dividing it by  
\begin{eqnarray}
\langle C_{i_0}^2(\tau)\rangle_\tau=
\frac{1}{2} \left(\frac{2}{N+1}\right)^2 \frac{N+1}{2}=
\frac{1}{N+1},
\label{ax2}
\end{eqnarray} 
one obtains the normalized secondary correlation 
\begin{eqnarray}
\!\!\!\!\!\!\!
D_{i_0}(t)=\frac{\langle C_{i_0}(\tau)\,C_{i_0}(\tau+t)\rangle_\tau}
{\langle C_{i_0}^2(\tau)\rangle_\tau}
=\frac{2}{N+1} 
\sideset{}{'}
\sum_{j=1}^N \cos\omega_j t
\end{eqnarray} 
which coincides with the primary correlation 
$C_{i_0}(t)$, Eq.(\ref{C}). 
 
\begin{figure}
  \includegraphics[height=6.4cm]{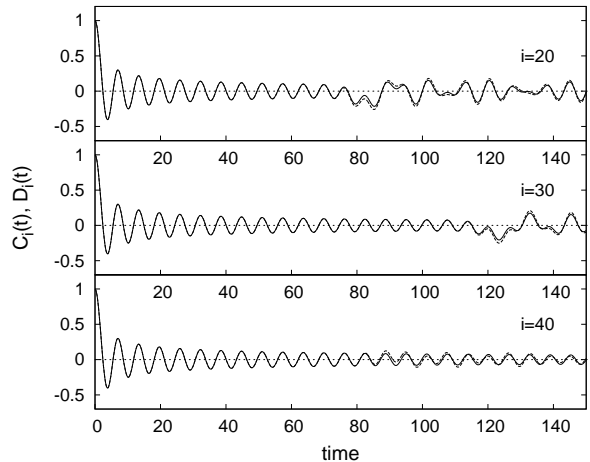}
\caption{The primary momentum correlation function $C_i(t)$ given by Eq. (\ref{C_gen}) (solid line)
and secondary correlation function $D_i(t)$ given by Eq. (\ref{D_gen}) (dashed line) 
for the harmonic chain with Hamiltonian (\ref{H}) with $N=101$ for
particles $i=20$ (top),
$i=30$ (middle), and $i=40$ (bottom).   The difference 
between $C_i(t)$ and $D_i(t)$ becomes noticeable for $t>t_0$ where $t_0$ 
depends on $i$ non-monotonically:
$t_0\approx 80, 120, 90$, from top to the bottom. 
 }
\label{fig_2}       
\end{figure}

One may observe that for the above derivation it is essential that 
the primary correlation $C_i(t)$ takes the form of a
superposition of cosines with equal weights, as in Eq. (\ref{C}).
In general this, of course, is not the case. For example, for 
a chain with fixed ends described by the Hamiltonian (\ref{H}),
the normalized momentum correlation function for particle with
arbitrary index
$i$ has the form (see Appendix)
\begin{eqnarray}
\!\!\!\!\!\!\!\!
C_i(t)=\sum_{j=1}^N A_{ij}^2\cos\omega_j t,\,\,\,\,\,
A_{ij}\!=\!\sqrt{\frac{2}{N+1}} \sin\frac{\pi ij}{N+1}.
\label{C_gen}
\end{eqnarray} 
For the middle particle $i=i_0=(N+1)/2$ this is reduced to (\ref{C}), whereas
for the other particles normal modes enter the expression 
(\ref{C_gen}) with different amplitudes $A_{ij}^2$. As one can immediately 
verify, the exact equality of primary and secondary 
correlations does not hold in these cases.
An important example when this  equality  does hold 
for {\it any} particle is a harmonic chain with periodic
boundary conditions. In this case the momentum correlation for each particle 
is a superposition of equally weighted normal 
modes~\cite{Mazur,Zwanzig}
\begin{eqnarray}
C_i(t)=\frac{1}{N}\sum_{j=0}^{N-1}\cos\omega_j t,\quad 
\omega_j=2\omega\sin\left(\frac{\pi j}{N}\right),
\label{periodic}
\end{eqnarray}
and repetition of the above derivation leads again to the exact equality 
$C_i(t)=D_i(t)$ for any particle of the system.

So far, even with the above  examples of its validity, the equality of primary and secondary correlations
may appear as no more than a curious coincidence. However, further 
numerical exercises reveal that even when equality does not hold exactly, 
it remains a very good approximation 
for the initial time interval $t<t_0$, see Fig. 2.       
The duration of this interval, $t_0$, is found to 
depend non-monotonically on particle position $i$, and 
for any $i$ be  equal or shorter
than the crossover time, $t_0\le t_c$. Respectively, 
for $t<t_0$  
both primary and secondary correlations coincide 
with the primary correlation for the infinite chain, 
\begin{eqnarray}
D_i(t)=C_i(t)=C_\infty(t)=J_0(2\omega t), \qquad t<t_0\le t_c.
\label{equality2}
\end{eqnarray}

%Broadly speaking, for a given $N$ 
%particles closer to the end of the chain will have a %smaller $t_0$.
%Yet remarkably, unlike $t_c$ the dependence of $t_0$ on %particle position $i$
%is not monotonic, even when considering particles from the
%same half of the chain (see Fig.~2). 

The proof of the approximate equality (\ref{equality2}) can be carried out 
as follows. From the expression (\ref{C_gen}) for $C_i(t)$ and the definition 
(\ref{D1}) for $D_i$, and using the relation (\ref{relation}),  one gets 
\begin{eqnarray}
D_i(t)=\frac{1}{\sum\limits_{j=1}^N A_{ij}^4}\,\,
\sum_{j=1}^N A_{ij}^4\,\cos\omega_j t,
\label{D_gen}
\end{eqnarray}
or taking into account the expression (\ref{omegas}) for normal 
mode frequencies 
\begin{eqnarray}
\!\!\!\!\!\!\!
D_i(t)=\frac{1}{\sum\limits_{j=1}^N A_{ij}^4}\,\,
\sum_{j=1}^N A_{ij}^4\,
\cos\Bigl[2\omega t\,\sin\left(
\frac{\pi j}{2(N+1)}\right)
\Bigr].
\end{eqnarray}
Recognizing here the generating function for Bessel functions
\begin{eqnarray}
\cos(x\sin\theta)=J_0(x)+2\sum_{k=1}^{\infty} J_{2k}(x)\,\cos(2k\theta),
\end{eqnarray}
$D_i(t)$  can be written as a superposition of Bessel functions
\begin{eqnarray}
D_i(t)=J_0(2\omega t)+\sum_{k=1}^{\infty} S_{ik}\,J_{2k}(2\omega t),
\label{D3}
\end{eqnarray}
with coefficients
\begin{eqnarray}
S_{ik}=\frac{2}{\sum\limits_{j=1}^N A_{ij}^4}\,\,
\sum_{j=1}^N A_{ij}^4\,
\cos\left(
\frac{\pi j k}{N+1}\right).
\label{S}
\end{eqnarray}
A simple  analysis of this expression  % LOL
shows that given $i$, the coefficients $S_{ik}$ are nonzero only for 
five sets of $k$:
\begin{eqnarray} 
S_{ik}=\begin{cases} \quad 2, &k=2(N+1)s\\
-4/3, &k=2(N+1)s- 2i  \\
-4/3, &k=2(N+1)(s-1)+2i \\
\quad 1/3, &k=2(N+1)s-4i \\
\quad 1/3, &k=2(N+1)(s-1)+4i\\
\quad 0,&\mbox{otherwise}
\end{cases} 
\label{S2}
\end{eqnarray}
where $s=1,2,3,\dots$ 
Note that this expression is invariant under the transformations 
$i\to (N+1)-i$, reflecting the symmetry of the left and right 
sides of the chain.
One can observe that for large $N$ and $i$ not too close to the end or to the middle of the chain the coefficients 
$S_{ik}$ are nonzero only for large indices $k$. 
For instance, for the chain with $N=101$ and the particle $i=20$, 
coefficients $S_{ik}$ are nonzero only 
for  $k=40, 80, 124,\dots$
As a result, for $t$ not too large in the expression 
(\ref{D3}), the dominating contribution comes from the first term 
$J_0(2\omega t)$,
while the corrections given by the sum 
$\sum_{k=1}^{\infty} S_{ik}\,J_{2k}(2\omega t)$  
involve Bessel functions of large orders which are negligibly small for a 
significant time interval $t<t_0$~\cite{Stegun}.

The above consideration not only 
justifies the equality $D_i(t)=C_i(t)=J_0(2\omega t)$ for $t<t_0$, 
but also accounts for a curious non-monotonic 
dependence of $t_0$ on the tagged particle index $i$,
which we noticed empirically in Fig.~2. 
For example, according to  (\ref{S2}), for $N=101$ and particles $i=20,30,40$ 
the minimal indices $k$ for which $S_{ik}$ takes nonzero values 
($-4/3, -4/3, 1/3$) are $k=40, 60, 44$, respectively.  
Then keeping only the leading and first correction terms
in the exact expression (\ref{D3}), one gets
\begin{eqnarray}
D_{20}(t)&=&J_0(2\omega t)-\frac{4}{3}\,J_{80}(2\omega t),\nonumber\\
D_{30}(t)&=&J_0(2\omega t)-\frac{4}{3}\,J_{120}(2\omega t),\nonumber\\
D_{40}(t)&=&J_0(2\omega t)+\frac{1}{3}\,J_{88}(2\omega t).
\label{correction}
\end{eqnarray}   
One can verify that these approximations describe
the initial deviation of $D_i(t)$ from $C_\infty(t)=J_0(2\omega t)$ 
very well indeed (Fig. 3 shows this for particle $i=30$ and $i=40$).
Since for small arguments $J_i(x)$ decreases with order 
$i$, it is clear from 
(\ref{correction}) that the second correction terms 
for particles $i=20, 40$ involve Bessel functions of smaller orders,
and thus become essential at earlier times than 
for particle $i=30$.

\begin{figure}
  \includegraphics[height=6.4cm]{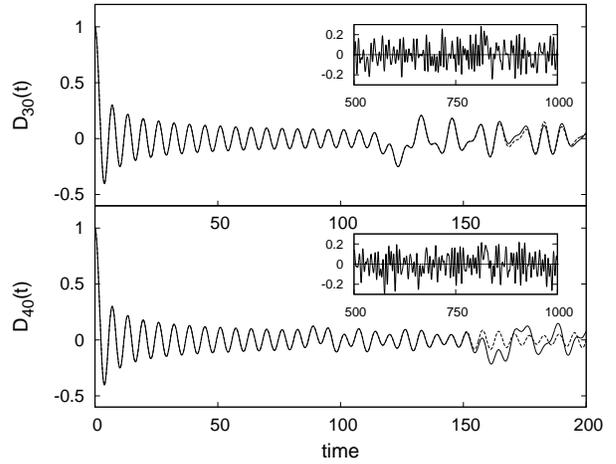}
  \caption{
The secondary correlation functions  $D_{30}(t)$ (top) 
and $D_{40}(t)$ (bottom)
according to the exact expression (\ref{D_gen})
(solid lines) and the approximation (\ref{correction})
(dashed lines). 
%For $t<150$ the approximation is very good for both functions. 
The insets show 
apparently random behavior of $D_{30}(t)$ and $D_{40}(t)$  at longer times. 
  }
  \label{fig_3}       % Give a unique label
\end{figure}

If one applies a similar analysis to the primary correlations $C_i(t)$,
Eq.~(\ref{C_gen}), one gets a familiar  approximate relation for the left side of the chain~\cite{Vig,Lee,PS} 
\begin{eqnarray}
C_i(t)=J_0(2\omega t)-J_{4i}(2\omega t).
\label{C5}
\end{eqnarray}
Here, in contrast to corresponding relations
(\ref{correction}) for $D_i(t)$, the order of the second Bessel function, 
which describes effects of finite size, increases monotonically (linearly) 
with particle index $i$, and so does the crossover time $t_c$.

In order to study the dependence of the characteristic time $t_0$, during which $C_i(t)=D_i(t)$, on
$i$ in a more quantitative way, let us consider the function
\begin{eqnarray}
\delta_i(t)=C_i(t)-D_i(t),
\label{delta}
\end{eqnarray}
which is zero when the two correlations coincide for $t<t_0$ and fluctuates at
longer times.  For a given $i$, let us define $t_0$ somewhat
arbitrarily as the time at which $\delta_i(t)$ reaches its first local
minimum or maximum; see Fig. 4(a).  
Similarly, we can define
the crossover time $t_c$ as the moment when the function
\begin{eqnarray}
\Delta_i(t)=C_i(t)-C_\infty(t)
\label{Delta}
\end{eqnarray}
has its first local extremum, recalling that
$C_\infty(t)=J_0(2\omega)$ is the correlation in an infinite system.
Using these definitions, we record observations of $t_0$ and $t_c$ for
$N=101$ in Fig.~4(b), as a function of particle index $i$.  Whereas $t_c$
increases linearly as we approach the central particle, $t_0$
coincides with $t_c$ for $i < i_1 = 34$ and linearly \emph{decreases}
for $i > i_1$. As we already know, the primary and secondary
correlations coincide for the middle particle $i_0=51$, so
$\delta_{i_0}(t)$ is identically $0$ and $t_0$ diverges here. Somewhat
unexpectedly, we find that $t_0$ also diverges, i.e. $C_i(t)=D_i(t)$ identically,  for $i_1=34$ (and of
course the symmetric case $i_2=(N+1)-i_1=68$). Therefore it would
appear that $t_0(i)$ diverges whenever it changes from increasing to
decreasing, or vice versa. Further calculations for different $N$ show that in general the exact
equality $C_i(t)=D_i(t)$ holds for particles with indices
\begin{eqnarray}
i_0=\frac{N+1}{2},\quad i_1=\frac{N+1}{3},\quad
i_2=\frac{2(N+1)}{3}
\label{special}
\end{eqnarray}
provided of course that these expressions are integers. For $N=101$
there are three such particles ($i_0=51, i_1=34, i_2=68$), two for $N=200$ ($i_1=67, i_2=134$), and none for $N=100$.

Let us show that this phenomenon is readily accounted for with
Eqs.(\ref{D3})-(\ref{S2}) for the secondary correlation $D_i(t)$.
First, from inspecting (\ref{S2}) one might observe that for $i<i_0$
the minimum $k$ for which $S_{ik}$ is non-zero is
$k=2i$ and comes from the set $k=2(N+1)(s-1)+2i$ with $s=1$. This
yields the approximation
\begin{eqnarray}
D_i(t)=J_0(2\omega t)-\frac{4}{3}\,J_{4i}(2\omega t),
\label{D5}
\end{eqnarray}
which we already used for $D_{20}(t)$ and $D_{30}(t)$ in
(\ref{correction}). It differs from the corresponding approximation
(\ref{C5}) for $C_i(t)$ only by the factor $4/3$ in the second term. 
Then, from (\ref{D5}) and (\ref{C5}),
the difference functions defined above by relations (\ref{delta}) and
(\ref{Delta}) take the form
\begin{eqnarray}
\delta_i(t)=\frac{1}{3}\,J_{4i}(2\omega t),\quad
\Delta_i(t)=-J_{4i}(2\omega t).
\label{deltas1}
\end{eqnarray}
Since these two functions have local extrema at the same time, by
definition we have $t_0=t_c$.  
Furthermore, since
the position of the first maximum 
of the Bessel function $J_i(t)$ increases approximately linear
with $i$~\cite{Watson},
Eq. (\ref{deltas1})  explains the
equality of the characteristic times $t_c(i)=t_0(i)$ and their linear
increase for $i<i_0$ in Fig. 4(b).

As $i$ gets larger still, one observes from (\ref{S2}) that a minimal
$k$ for which $S_{ik}\ne 0$ is $k=2(N+1)-4i$ and comes from the set
$k=2(N+1)s-4i$ with $s=1$.  In this case for $D_i(t)$, instead of
(\ref{D5}), we have another approximation
\begin{eqnarray}
D_i(t)&=&J_0(2\omega t)+\frac{1}{3}\,J_{\alpha}(2\omega t),\nonumber\\
\alpha&=&4(N+1)-8i,
\label{D6}
\end{eqnarray}
which we already used for $D_{40}(t)$ in (\ref{correction}). Since the
primary correlation $C_i(t)$ is still given by (\ref{C5}), the
difference function $\delta_i(t)=C_i(t)-D_i(t)$ in this case reads
\begin{eqnarray}
\delta_i(t)=-\frac{1}{3}\,J_{\alpha}(2\omega t)-J_{4i}(2\omega t)\approx -\frac{1}{3}\,J_{\alpha}(2\omega t).
\end{eqnarray}
The position of its first extremum increases approximately linearly
with $\alpha$~\cite{Watson} and, as follows from (\ref{D6}), linearly decreases with
$i$. This explains the behavior of $t_0(i)$ for $i>i_1$ in Fig. 4(b).

\begin{figure}
  \includegraphics[height=6cm]{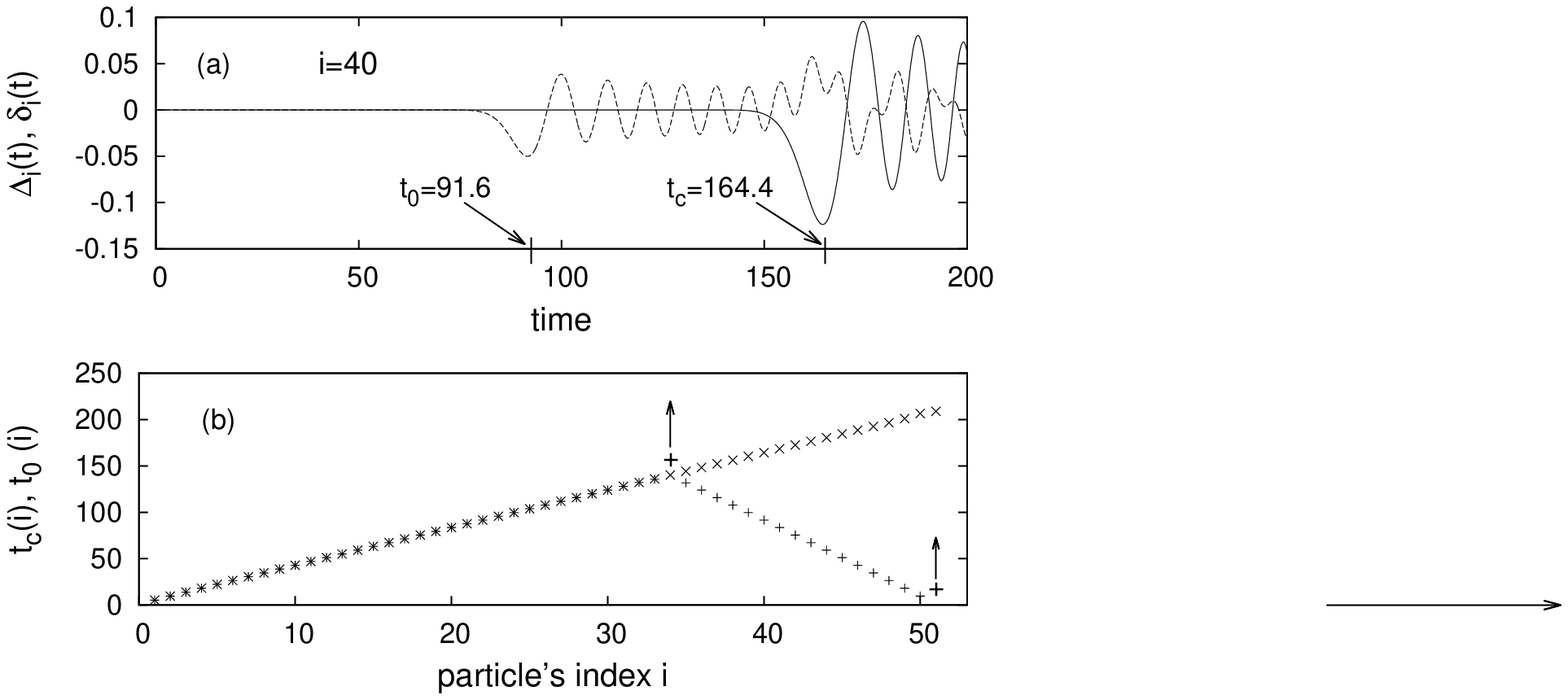}
  \caption{
  Top plot (a): The functions $\Delta_i(t)=C_i(t)-J_0(2\omega t)$ (solid line) and
    $\delta_i(t)=C_i(t)-D_i(t)$ (dashed line) for particle
    $i=40$. The characteristic times $t_c$ and $t_0$ are defined as
    times at which $\Delta_i(t)$ and $\delta_i(t)$, respectively, have
    their first local extremum.
    Bottom plot (b): The characteristic times $t_c$ ($\times$) and $t_0$ ($+$)
    for particles with indices $i\le 51$ for the left side of the
    chain described by Hamiltonian (\ref{H}) with $N=101$. For
    particles $i_0=51$ and $i_1=34$ the time $t_0$ diverges (illustrated by an arrow 
    pointing upward), indicating the exact equality $C_i(t)=D_i(t)$. 
    }
\label{fig_4}       
\end{figure}

The transition of $t_0(i)$ from a positive to a negative slope (over the 
domain $[0,51]$) occurs at $i=i_1$, for which $k_2=2(N+1)-4i$ (the minimal value of the set $k=2(N+1)s-4i$) becomes less than or equal to $k_1=2i$ (the minimal value of the set $k=2(N+1)(s-1)+2i$). Then the equality $k_1=k_2$ gives $i_1=(N+1)/3$,
which is consistent with our empirical findings (\ref{special}).

The exact equality $C_i(t)=D_i(t)$ for $i$ given by (\ref{special})
can be readily verified using the following expression for the primary
correlations
\begin{eqnarray}
C_i(t)=J_0(2\omega t)+\sum_{k=1}^{\infty} T_{ik}\,J_{2k}(2\omega t),
\label{C7}
\end{eqnarray}
with coefficients
\begin{eqnarray}
T_{ik}=2\,
\sum_{j=1}^N A_{ij}^2\,
\cos\left(
\frac{\pi j k}{N+1}\right).
\label{T}
\end{eqnarray}
These relations are similar to (\ref{D3}) and (\ref{S}) for $D_i(t)$ and
can be derived in a similar way~\cite{PS}. For $i=i_0, i_1, i_2$ given
by (\ref{special}), one can verify directly from (\ref{T}) and (\ref{S}) that
$S_{ik}=T_{ik}$ for any $k$. Then the comparison of (\ref{C7}) and
(\ref{D3}) gives for those values of $i$ the exact equality
$C_{i}(t)=D_{i}(t)$.

\section{Heavy impurity problem}
So far we have discussed finite harmonic systems of similar particles.
If a tagged particle is heavier 
than the rest, it turns out that the equality of
primary and secondary correlations, $C(t)$ and $D(t)$, does not hold. Though structurally similar - $D(t)$ looks like a coarse-grained copy of $C(t)$ - the two correlations are quite distinctive on any time scale; see Fig. 5.  In particular, the approximation of exponential relaxation for $t<t_c$, while good for $C(t)$, is noticeably worse for $D(t)$. Another observation is that for $t\gg t_c$ both correlations, being apparently random on a short time scale, show  
on a larger scale a noisy yet periodically repeating pattern; see the bottom plot in Fig. 5. This feature, absent 
in systems of equal-mass particles, 
is made all the more obvious when considering higher order 
correlation functions $C_k(t)$, defined recursively as 
\begin{eqnarray}
C_{k+1}(t)=\frac{\langle C_k(\tau)C_k(\tau+t)\rangle_\tau}
{\langle C_k^2(\tau)\rangle_\tau},
\label{CCC}
\end{eqnarray}
assuming new notations for $C(t)=C_1(t)$ and $D(t)=C_2(t)$.
(In this section we use the notation
$C_k(t)$ with a subscript referring to the correlation order, rather than to the index of a particle). 
For the heavy impurity problem one finds that as the order $k$ increases the apparent randomness of correlations $C_k(t)$ on 
the time scale $t>\tau_c$ quickly \emph{diminishes}, and  
higher correlations converge 
to the normal mode with the lowest eigenfrequency $\Omega_1$:   
\begin{eqnarray}
C_k(t)\to \cos(\Omega_1 t);
\label{apple}
\end{eqnarray}
see Fig. 6. Below we outline a theoretical framework underlying these empirical observations.

\begin{figure}
  \includegraphics[height=6.4cm]{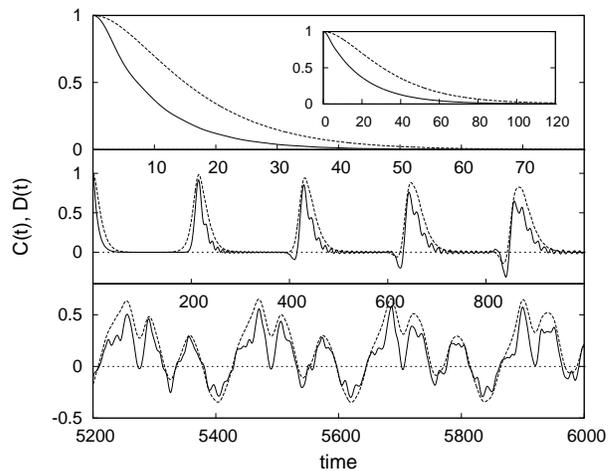}
\caption{
The primary (solid line) and secondary (dashed line) momentum correlation functions, $C(t)$ and $D(t)$, for the heavy impurity problem described by Hamiltonian (\ref{HH}) with mass ratio $\mu=m/M=0.1$ and $N=50$. 
The top, middle, and bottom figures show the evolution of the correlations on short ($t< t_c$), intermediate ($t\sim t_c$), and long ($t\gg t_c$) time scales. The inset shows correlations on the short time scale for an impurity that is twice as heavy; $\mu=0.05$. Time is in units of $1/2\omega$.
}
\label{fig_5}       
\end{figure}

Consider a cyclic chain of $2N$ particles 
of mass $m$ and an
impurity of mass $M>m$ described by the Hamiltonian
% P and Q are the momentum and coordinates corresponding to the imprity?
\begin{eqnarray}
H&=&\frac{P^2}{2M}+\sum_{i=1}^{2N}\frac{p_i^2}{2m}
+\frac{m\omega^2}{2}\sum_{i=1}^{2N-1}(q_i-q_{i+1})^2\nonumber\\
&+&\frac{m\omega^2}{2}\Large[(Q-q_1)^2+(Q-q_{2N})^2\Large],
\label{HH}
\end{eqnarray} 
where $P$ and $Q$ are the momentum and coordinates of the impurity.
Using a diagonalization method similar to that described in the Appendix
(see \cite{Cukier} for details),
one can show that the normalized momentum correlation function for the impurity 
$C(t)=\langle P(0)P(t)\rangle/\langle P^2(0)\rangle$ 
is again an almost periodic function,  now of the form
\begin{eqnarray}
C(t)\equiv C_1(t)=  
\sum_{j=0,1,3,\cdots}^{2N-1}A_j\,\cos \Omega_j t.
  \label{CC}
\end{eqnarray}
The amplitudes $A_j$ in this expression are given by
\begin{eqnarray}
A_j=\left\{1+\sum_{i=1,3,\cdots}^{2N-1}\,\frac{\epsilon_i^2}{(\Omega_j^2-\omega_i^2)^2}
\right\}^{-1}.
\label{AA}
\end{eqnarray}
where
\begin{eqnarray}
\omega_i&=&2\omega\, \sin\frac{i\,\pi}{2(2N+1)},\nonumber\\
\epsilon_i&=&-2\mu^{\frac{1}{2}}\,\omega^2\,\left(
\frac{2}{2N+1}\right)^\frac{1}{2}
\,\sin\frac{i\,\pi}{2N+1},
\end{eqnarray}
and $\mu=m/M$ is the mass ratio.

Due to the system's symmetry only the modes with zero and odd indices 
contribute to the superposition (\ref{CC}). 
Their frequencies
$\Omega_j$ ($j=0,1,3,\cdots 2N-1)$
for $M\ne m$ cannot be expressed in closed form 
and must be evaluated   
as roots of the secular equation~\cite{Cukier}
\begin{eqnarray}
G(z)=z^2-2\,\mu\,\omega^2-
\sum_{i=1,3,\cdots}^{2N-1}\,\frac{\epsilon_{i}^2}{z^2-\omega_{i}^2}=0.
\label{secular}
\end{eqnarray}
This transcendental equation has  $N+1$ solutions 
$z=\Omega_j$, $j=0,1,3,\cdots 2N-1$.
It can be verified that one solution is the zero frequency $\Omega_0=0$,
which reflects the translational invariance of the system.
The remaining $N$ nonzero roots $\Omega_1, \Omega_3,\dots\Omega_{2N-1}$
lie in the interval $(0,2\omega)$ and must be evaluated numerically.

\begin{figure}
  \includegraphics[height=6.0cm]{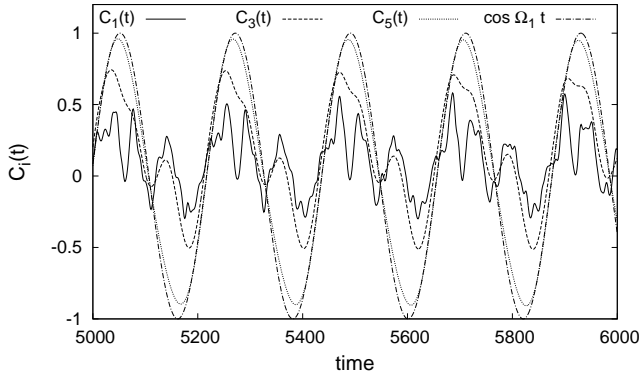}
  \caption{
The primary correlation $C_1(t)$ (solid line), and two higher correlations  $C_3(t)$ (dashed line) and $C_5(t)$ (dotted line), defined by Eq. (\ref{CCC}), at long time $t\gg t_c$ for the heavy impurity problem with $\mu=0.1$ and $N=50$. 
Higher correlations converge to  the lowest normal mode $\cos\Omega_1 t$ (dot dashed line). 
  }
  \label{fig_6}      
\end{figure}

With the set of eigenfrequencies $\Omega_j$ found, one may calculate
the amplitudes $A_j$ with (\ref{AA}) 
and evaluate the primary correlation $C(t)$ 
by carrying out the summation in (\ref{CC}). Then, using
(\ref{relation}), for the  secondary correlation $D(t)$ one obtains
\begin{eqnarray}
D(t)=C_2(t)=c_2\,\left\{A_0^2+\frac{1}{2}
\sum\limits_{j=1,3,\cdots}^{2N-1}
A_{j}^2\,\cos\Omega_j t\right\}
\label{DD}
\end{eqnarray}
with normalization coefficient
\begin{eqnarray}
c_2=\left(
A_0^2+\frac{1}{2}\sum\limits_{j=1,3,\cdots}^{2N-1} A_{j}^2
\right)^{-1}.
\label{DD0}
\end{eqnarray}
Fig. 5 presents $C(t)$ and $D(t)$, calculated with Eqs. (\ref{CC}) and (\ref{DD}),
for $\mu=0.1$ and $N=50$. 
In a similar manner, one can obtain
the expression for order-$k$ correlations from (\ref{CCC})
\begin{eqnarray}
C_k(t)=c_k\,\left\{A_0^{\alpha_k}+
\sum\limits_{j=1,3,\cdots}^{2N-1}
\left(\frac{A_{j}}{\sqrt{2}}\right)^{\alpha_k}\,\cos\Omega_j t\right\}
\label{CCCC}
\end{eqnarray}
with powers $\alpha_k=2^{k-1}$ and normalization coefficient
\begin{eqnarray}
c_k=\left\{A_0^{\alpha_k}+
\sum\limits_{j=1,3,\cdots}^{2N-1}
\left(\frac{A_{j}}{\sqrt{2}}\right)^{\alpha_k}\right\}^{-1}.
\end{eqnarray}
While expression (\ref{CCCC}) is a superposition of $N+1$ modes,
one can observe that for larger $k$ 
the main contribution comes from the mode with eigenfrequency $\Omega_1$,
such that $C_k(t)\approx\cos\Omega_1 t$. 
This can be accounted for by noticing that 
the sequence of coefficients $\{A_j\}$ has $A_1$ as its maximum
element and is monotonically decreasing for $j>0$. For example, 
for $\mu=0.1$ and $N=50$ we find that
$A_0=0.09, A_1=0.17, A_2=0.15, A_3=0.12,\dots$ (approximately). 
For the primary and secondary correlations involving $A_j$ and $A_j^2$
respectively, such an insignificant difference in values hardly plays a
role. But for higher-order correlations the maximum of the set 
$\{A_j^{\alpha_k}\}$ (still at $j=1$) may be orders of magnitude greater
than any other element. As a result, the superposition in (\ref{CCCC}) is
increasingly dominated by the term with $A_1^{\alpha_k}$, and higher-order
correlations quickly
converge to the first normal mode; $C_k(t)\approx\cos\Omega_1 t$.

For systems of same-mass particles   
the set of normal mode amplitudes, given by the second equation in (\ref{C_gen}), is a periodic function of the mode index $j$ and has no single maximum. In this case the reduction of higher-order correlations to a dominating normal mode does not occur.

\section{Conclusion}
Temporal autocorrelation functions $\langle A(0)A(t)\rangle$  
are often evaluated in the thermodynamic limit,
in which case they typically decrease in a regular (non-random) fashion,
either monotonically or non-monotonically. 
% Broadly speaking, they characterize how long a given value of
% a noisy dynamical variable $A$ lasts on average. 
In finite systems, autocorrelation functions themselves
become noisy at long time scales $t>t_c$; this illustrates
recurrences in the dynamics of the tagged variable due to reflections of
sound off boundaries.
In this paper we introduced and studied some properties of 
the secondary correlation function $D(t)$ defined as an autocorrelation
function of the primary correlation function $C(t)$.
If it exists, the characteristic time of decay for
$D(t)$ determines the time-scale of a typical ``period'' for $C(t)$, 
which in turn may be associated with the typical recurrence time of 
the targeted variable. 
These ``typical'' times may however be ill-defined mathematically 
(as is indeed the case for the harmonic systems discussed above), so to be
more precise the secondary correlation $D(t)$ can be described as a function 
characterizing a {\it distribution} of recurrence times:
for a given $t$, a larger value for $D(t)$ corresponds to a greater
probability (density) that $C(t)$ will return to an assigned value in time 
about $t$. 
Comparing the secondary correlation $D(t)$ with the mean recurrence time
$\tau(c)$, Eq.~(\ref{Kac}), the latter being more prevalent in literature, 
one notices that the two functions give complementary descriptions:
while $\tau(c)$ characterizes the number of returns to an assigned value
$c$, the secondary correlation $D(t)$ gives the 
distribution of return times regardless of the assigned value of $c$.

One interesting result is the equality $C(t)=D(t)$ 
for systems of same-mass particles.
The equality  
is either exact for all $t$ or a very good approximation over 
the initial interval $t<t_0$ whose duration $t_0$ depends on the tagged 
particle's position 
non-monotonically. Although its derivation is quite simple, 
the equality of primary and secondary correlations 
may be a remarkable property,
especially considering that the former is defined over the ensemble
and the latter with time averaging. 
We restricted the discussion to the simplest case of one-dimensional
harmonic systems, but  an extension to higher dimensions appears to be
straightforward. Whether the equality, or perhaps some other relation, 
between primary and secondary correlations still holds for nonlinear systems
is an open question.

Like the primary correlation, for long time-scales 
the secondary correlations also develop noisy tails 
(see the insets in Fig. 3),
which themselves can be characterized by correlations of higher order. 
In turn, this new tertiary function has the same structure as the 
secondary and primary functions,
exhibiting regular decay over shorter times and fluctuating over 
longer times. Thus one can construct an infinite hierarchy of 
higher order correlations whose
order-scaling properties are also interesting to study. 
Of course, for the particular cases
when the equality $D(t)=C(t)$ holds exactly for all $t$, 
e.g.~a harmonic chain with periodic boundary conditions, all higher order correlations are identical. 
We have studied higher correlations in the context of the heavy impurity problem. Here the equality of primary and secondary correlations does not hold, and the primarily correlation displays a non-random idiomatic pattern on long time scales, which becomes even more visible in correlations of higher orders. Indeed the sequence of higher-order correlations converges to 
 the lowest normal mode.

\begin{acknowledgements}
We thank S. Shea and V. Dudnik for discussion and the anonymous referee for  important suggestions.
\end{acknowledgements}

\section*{Appendix}
In this appendix we outline the derivation of
expressions (\ref{C}) and (\ref{C_gen}) for the momentum correlation functions of $i$-th particle 
in a harmonic chain with fixed ends, described by the Hamiltonian 
(\ref{H}).
Using the normal mode transformation
\begin{eqnarray}
q_i=\frac{1}{\sqrt{m}}\sum_{j=1}^N A_{ij} Q_j,\quad
p_i=\sqrt{m}\sum_{j=1}^N A_{ij} P_j
\nonumber
\end{eqnarray}
with coefficients $A_{ij}$ given by (\ref{C_gen}) 
and taking into account the orthogonality relation
$\sum_{i=1}^N A_{ij}A_{ij'}=\delta_{jj'}$, the Hamiltonian (\ref{H}) 
is diagonalized into the form of uncoupled normal modes
\begin{eqnarray}
H=\frac{1}{2}\sum_{j=1}^N \{P_j^2+\omega_j^2Q_j^2\}
\nonumber
\end{eqnarray}
with frequencies $\omega_j$ given by (\ref{omegas}).
The normal modes are governed by the Hamiltonian equations
\begin{eqnarray}
\dot P_j=-\frac{\partial H}{\partial Q_j}=-\omega_j^2 Q_j,\quad
\dot Q_j=\frac{\partial H}{\partial P_j}=P_j,\quad
\nonumber
\end{eqnarray}
and evolve as
\begin{eqnarray}
P_j(t)&=&P_j(0)\cos\omega_j t-\omega_jQ_j(0)\sin\omega_jt,\nonumber\\
Q_j(t)&=&Q_j(0)\cos\omega_j t+\omega_j^{-1}P_j(0)\sin\omega_jt.
\nonumber
\end{eqnarray} 
Assuming that initially the system is in equilibrium with canonical 
distribution function $\rho_e=Z^{-1}e^{-\beta H}$, correlations of
the normal modes' initial values are
$\langle P_j(0)P_{j'}(0)\rangle=\delta_{jj'}/\beta$ and 
$\langle Q_j(0)P_{j'}(0)\rangle=0$.
%\langle Q_j(0)Q_{j'}(0)\rangle=\frac{\delta_{ij}}{\beta\omega_j^2}
Then
\begin{eqnarray}
\langle P_{j'}(0)P_{j}(t)\rangle=
\langle P_{j'}(0)P_{j}(0)\rangle\,\cos\omega_j t=
\frac{\delta_{jj'}}{\beta}\,\cos\omega_jt\nonumber
\end{eqnarray}
and the momentum correlation of the $i$-th particle is
 \begin{eqnarray}
\langle p_i(0)\,p_i(t)\rangle&=&
m\,\sum_{j,j'=1}^N \!\!A_{ij}A_{ij'}\langle
P_{j'}(0)P_{j}(t)\rangle\nonumber\\
&=&\frac{m}{\beta}\, \sum_{j=1}^N \!A_{ij}^2 \cos\omega_j t.
\nonumber
\end{eqnarray}
Division of this expression by $\langle p_i^2(0)\rangle=m/\beta$ 
gives the normalized correlation
function (\ref{C_gen}). 
In the case of the middle particle $i=(N+1)/2$ (assuming $N$ is odd),
$A_{ij}^2 = 2/(N+1)$ for odd $j$ and zero otherwise.
In this case, one obtains the normalized correlation function
in the form (\ref{C}). 
The correlation (\ref{periodic}) 
corresponding to the periodic boundary condition can be derived in a similar 
way. 
For the extension to the heavy impurity problem 
see e.g.~\cite{Cukier}.

\end{document}